\documentclass[12pt,preprint]{aastex}
\usepackage{emulateapj5}
\usepackage{epsfig}
                                                                                
\def\vkm{km s$^{-1}$}

\def\degree{$^\circ$}
\def\arcs#1{$#1''$}
\def\arcsa#1#2{$#1^{\prime\prime}_{^\textrm{.}}#2$}
\def\arcsaq#1#2{#1^{\prime\prime}_{^\textrm{.}}#2}
\def\solarmass{$M_\odot$}

\def\Jyb{Jy beam$^{-1}$}
\def\mJyb{mJy beam$^{-1}$}
\def\Jybk{Jy beam$^{-1}$ km s$^{-1}$}

\def\tlabel#1{(\textit{#1})}

\def\Volker{V$\ddot{\textrm{o}}$lker}
\def\cmc{cm$^{-3}$}
\def\cms{cm$^{-2}$}

\def\Vsys{V_\textrm{\scriptsize sys}}

\def\ra#1#2#3#4{#1^\mathrm{h} #2^\mathrm{m} #3^\mathrm{s}_{^\textrm{.}} #4}
\def\dec#1#2#3#4{#1\degr #2\arcmin #3^{\prime\prime}_{^\textrm{.}}#4}

\def\mH2{m_{\textrm{\scriptsize H}_2}}

\def\H2{H$_2$}
\def\N2HP{N$_2$H$^+$}
\def\HCOP{HCO$^+$}

\def\NH3{NH$_3$}

\def\HCOP{HCO$^+$}

\def\putfiga#1#2#3{\epsfig{scale=#1,angle=#2,figure=#3}}
\def\putfig#1#2#3{}
\def\leftblank#1{}
\def\bf#1{}

\begin{document}

\title{Jet Motion, Internal Working Surfaces, and Nested Shells in the
Protostellar System HH 212}

\author{Chin-Fei Lee\altaffilmark{1}, Naomi Hirano\altaffilmark{1},
Qizhou Zhang\altaffilmark{2}, Hsien Shang\altaffilmark{1}, 
Paul T.P. Ho\altaffilmark{1,2}, and Yosuke Mizuno\altaffilmark{3}
}
\altaffiltext{1}{Academia Sinica Institute of Astronomy and Astrophysics,
P.O. Box 23-141, Taipei 106, Taiwan; cflee@asiaa.sinica.edu.tw}
\altaffiltext{2}{Harvard-Smithsonian Center for Astrophysics, 60 Garden
Street, Cambridge, MA 02138}
\altaffiltext{3}{Institute for Theoretical Physics,
Frankfurt am Main, 60438, Germany}

\begin{abstract}

HH 212 is a nearby (400 pc) highly collimated protostellar jet powered by a
Class 0 source in Orion.  We have mapped the inner \arcs{80} ($\sim$ 0.16
pc) of the jet in SiO ($J=8-7$) and CO ($J=3-2$) simultaneously at $\sim$
\arcsa{0}{5} resolution with the Atacama Millimeter/Submillimeter Array
at unprecedented sensitivity.  The jet consists of a chain of knots, bow
shocks, and sinuous structures in between.  As compared to that seen in our
previous observations with the Submillimeter Array, it appears to be more
continuous, especially in the northern part.  Some of the knots are now seen
associated with small bow shocks, with their bow wings curving back to the
jet axis, as seen in pulsed jet simulations.  Two of them are reasonably
resolved, showing kinematics consistent with sideways ejection, {\bf
possibly} tracing the internal working surfaces formed by a temporal
variation in the jet velocity.  In addition, nested shells are seen in CO
around the jet axis connecting to the knots and bow shocks, driven by them.
The proper motion of the jet is estimated to be $\sim$ 115$\pm$50 \vkm{},
comparing to our previous observations.  The jet has a small semi-periodical
wiggle, with a period of $\sim$ 93 yrs.  The amplitude of the wiggle
first increases with the distance from the central source and then stays
roughly constant.  One possible origin of the wiggle could be the kink
instability in a magnetized jet.

\end{abstract}

\keywords{stars: formation --- ISM: individual: HH 212 --- 
-- ISM: jets and outflows.}

\section{Introduction}

Protostellar jets represent one of the most intriguing signposts of star
formation \cite[for recent review see][]{Frank2014}.  They are highly
supersonic and collimated, and some with wiggles in their trajectories. 
They are seen with knotty and bow-like shock structures.  They can be
launched from accretion disks around protostars, allowing us to probe the
accretion process closest to the protostars.  Two competing yet similar MHD
models, the X-wind model \citep{Shu2000} and disk-wind model
\citep{Konigl2000}, have been proposed to launch these jets.  Preliminary
measurements of specific angular momentum of some jets are consistent with
these two models \cite[see, e.g.,][]{Lee2008,Coffey2011}.  Polarization
observations also suggest that the jets could be magnetized
\citep{Carrasco2010,Lee2014b}, as predicted in the two models.  Therefore,
more detailed studies of the jet properties are needed to determine the jet
launching model, and thus the accretion process closest to the protostars.

Molecular outflows are shell-like structures surrounding the jets and thus
can be used to probe the physical properties of the jets indirectly.  They
are believed to be outflow cavity walls consisting mainly of the ambient
material swept-up by the bow shocks in the jets \cite[see,
e.g.,][]{Lee2001}.  Since some molecular outflows show wide-opening cavity
walls near the protostars, especially in the later stage of star formation
\citep{Lee2005,Arce2013}, an additional unseen wide-angle tenuous winds
might be required.  Indeed, current theoretical jet models also predict such
winds surrounding the highly collimated jets \citep{Shu2000,Konigl2000}. 
However, more observations are still needed to confirm this possibility,
especially in the early phase of star formation.

This paper is a follow-up study to our previous Submillimeter Array (SMA)
study of the HH 212 jet and outflow system \citep{Lee2007,Lee2008},
presenting our study in SiO and CO with Atacama Millimeter/Submillimeter
Array (ALMA) at an angular resolution of $\sim$ \arcsa{0}{5} with more than
10 times higher sensitivity.  HH 212 system is nearby in the L1630 cloud of
Orion at a distance of $\sim$ 400 pc.  Its jet was discovered in shock
excited \H2{} emission \citep{Zinnecker1998}.  It is powered by a
low-luminosity ($\sim$ 9 $L_\odot$) Class 0 protostar IRAS 05413-0104.  It
interacts with the ambient material, driving a collimated CO outflow around
it \citep{Lee2000}.  A rotationally supported disk is believed to have
formed around the protostar in order to launch the jet
\citep{Lee2014,Codella2014}.  Lying close to the plane of the sky
\cite[$\lesssim$ 5\degree{},][]{Claussen1998,Lee2007}, the jet is one of the best
candidates to investigate the jet properties.  Here, we refine the proper
motion and thus the mass-loss rate of the jet.  We study the shock
structures in the jet, the morphological relationship between the CO outflow
and the jet, and discuss if there is a need of a wide-angle tenuous wind
around the jet.  The jet has a small semi-periodical wiggle.  We model the
wiggle and discuss possible origins for the wiggle.  Note that the small jet
rotation tentatively found at higher resolution of $\sim$ \arcsa{0}{35} in
SiO \citep{Lee2008} can not be confirmed here at a lower resolution of
$\sim$ \arcsa{0}{5}.

\section{Observations}\label{sec:obs}

Observations of the HH 212 protostellar system were carried out with ALMA on
2012 December 1 during the Early Science Cycle 0 phase.  The details of
these observations have been reported in \citet{Lee2014}, and thus only
important information is reported here.  A 9-pointing mosaic was used to
observe the jet in this system within $\sim$ \arcs{40} from the central
source.  Both CO J=3-2 and SiO J=8-7 lines were observed at a velocity
resolution of $\sim$ 0.21 \vkm{} per channel.  With super-uniform weighting,
the synthesized beam has a size of \arcsa{0}{53}$\times$\arcsa{0}{47} at a
position angle (P.A.) of $\sim$ 20\degree{} in SiO, and
\arcsa{0}{56}$\times$\arcsa{0}{47} at a P.A.  of 43\degree{} in CO.  The rms
noise level is $\sim$ 7.7 \mJyb{} (i.e., 0.32 K) for the SiO channel maps,
and $\sim$ 7.2 \mJyb{} (i.e., 0.28 K) for the CO channel maps.  The
velocities in the channel maps are LSR.  The systemic velocity in this
region is assumed to be $\Vsys= 1.7\pm0.1$ \vkm{} LSR, as in
\citet{Lee2007}.  The central source is assumed to be at the peak position
of the continuum, which is $\alpha_{(2000)}=\ra{05}{43}{51}{4107}$,
$\delta_{(2000)}=\dec{-01}{02}{53}{167}$, as in \citet{Lee2014}.

\section{Results}

In order to provide readers with a more complete picture of the jet and
outflow system, our ALMA maps are plotted together with an earlier H$_2$ map
of the jet.  Also, for a better view, all the maps are rotated by
22.5\degree{} clockwise so that the jet axis is aligned with the y-axis. 
The H$_2$ map shows that the jet consists of a chain of knots and bow
shocks, and sinuous (continuous) structures in between.  It was made by
combining the 3 observations in October 2000, October 2001, and January 2002
\citep{McCaughrean2002}, and is thus roughly 11 years earlier than our ALMA
observations.  Since the proper motion of the jet is estimated to be $\sim$
\arcsa{0}{06} per year later here in our paper, the \H2{} jet must have
moved down the jet axis by $\sim$ \arcsa{0}{7}, as compared to the jet seen
in our ALMA maps.  Since this position shift is smaller than the size of a
knot ($\gtrsim$ \arcs{1}), the \H2{} map can still be used to show the rough
positions of the knots, bow shocks, and sinuous structures to be compared
with our ALMA maps.

\subsection{Nested Shells in CO}

\begin{figure} [!hbp]
\centering
\putfiga{0.75}{0}{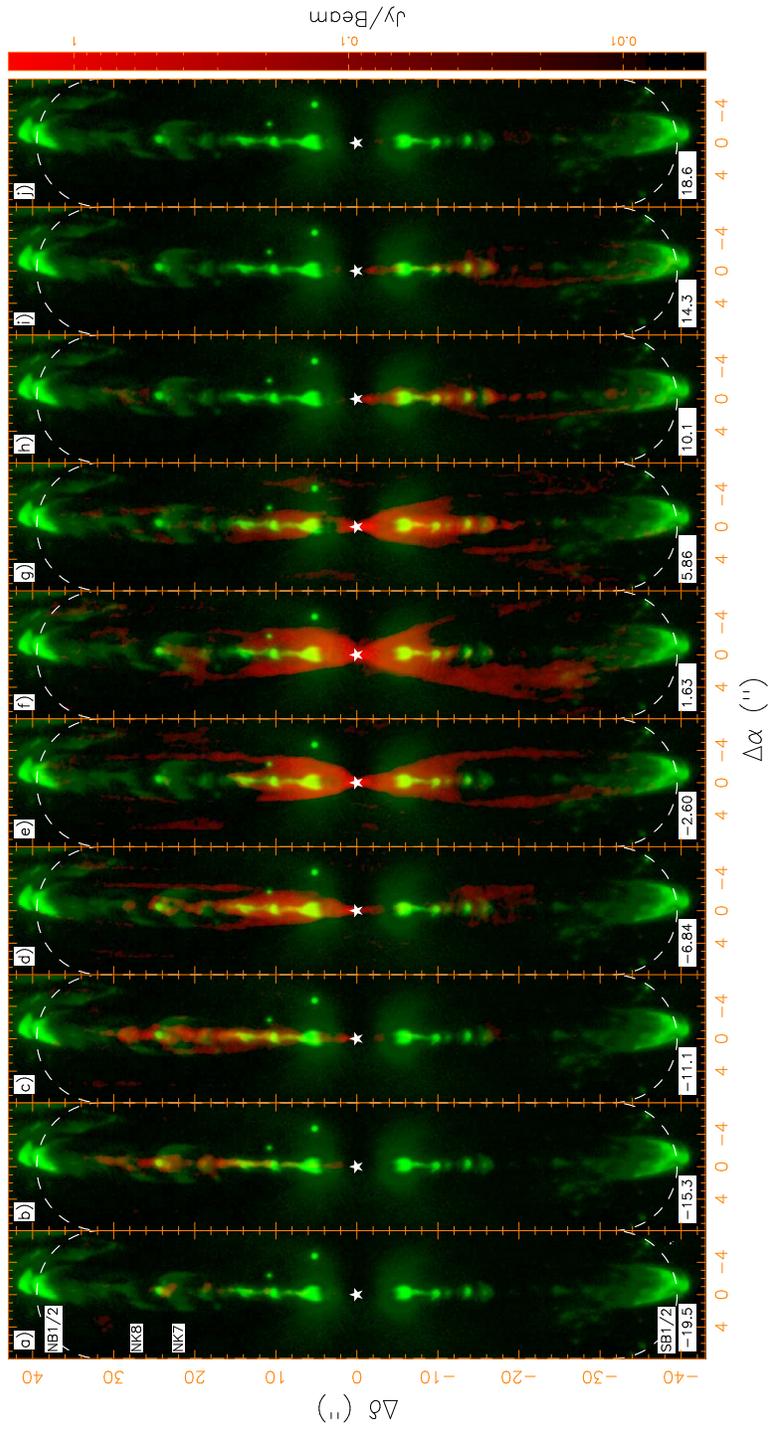} 
\figcaption[]
{CO channel maps (Red, in logarithmic scale) in comparison with the \H2{}
map \cite[Green, adopted from][]{McCaughrean2002} of the HH 212 system.  The
star marks the central source position.  The channel velocity is indicated
in the lower left corner.  The dashed arcs outline the primary beam coverage
in our ALMA observations.  The synthesized beam in the CO maps has a size of
\arcsa{0}{56}$\times$\arcsa{0}{47} with a PA $\sim$ 20\degree{}, 
as shown in the lower right corner in \tlabel{a}.
\label{fig:CO_chan}}
\end{figure}

\begin{figure} [!hbp]
\centering
\putfiga{1.0}{270}{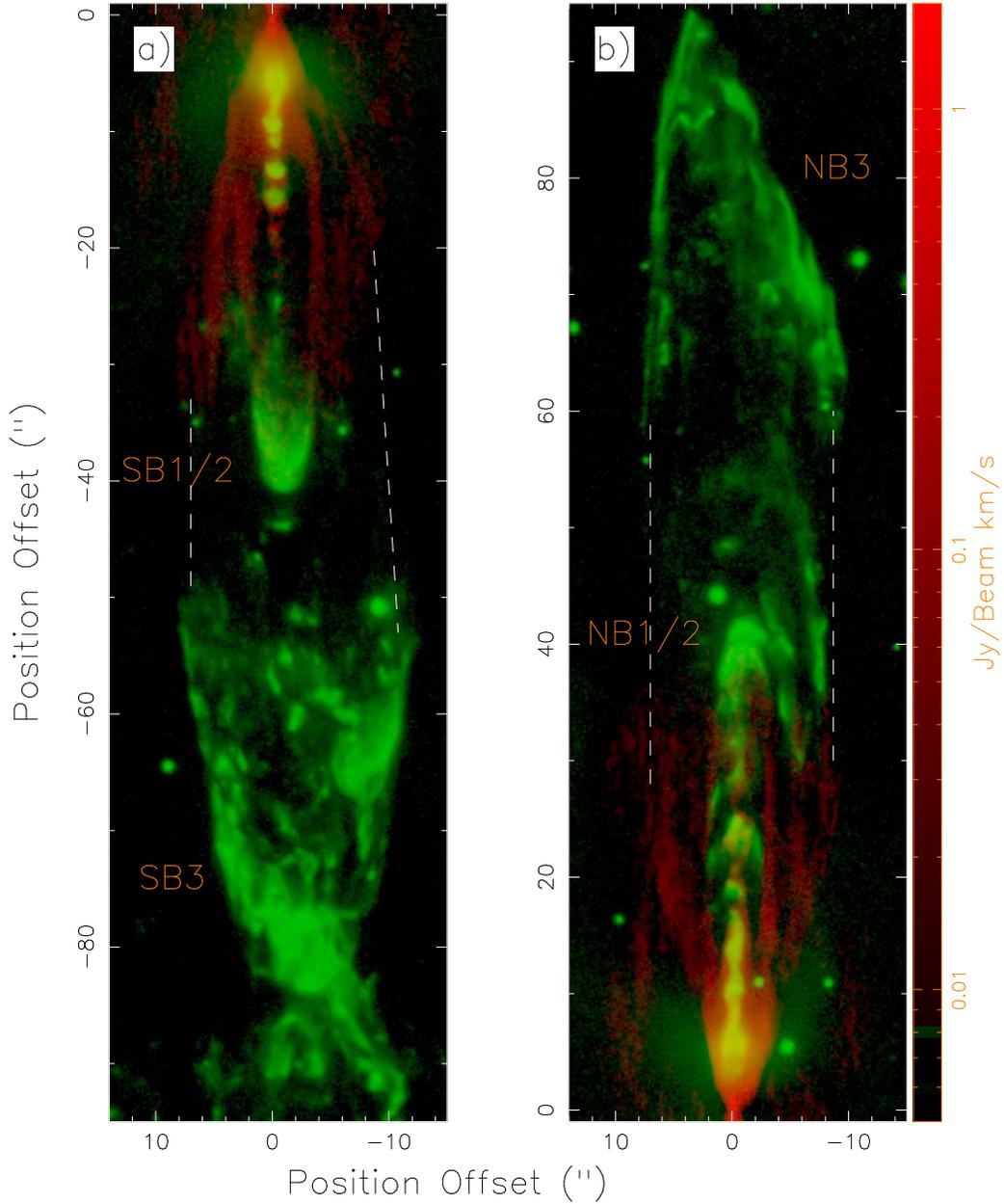} 
\figcaption[]
{CO maps (red) integrated over selected velocity ranges that show the shell
structures, in comparison to the \H2{} map (green).  In \tlabel{a}, the
velocities are from $-$3.6 to $-$2.9 \vkm{}, from $-$2 to $-$1.3 \vkm{}, from 1.7 to
2.9 \vkm{}, and from 4.9 to 6.1 \vkm{}.  In \tlabel{b}, the velocities are
from $-$7.8 to $-$5.6 \vkm{}, from $-$3.6 to 0.25 \vkm{}, and from 1.6 to 1.8
\vkm{}.  Nested CO shells are seen connecting to the \H2{} bow shocks and
knots.  The dashed lines indicate the connections of the wide-opening shells
to the big bow shocks, SB3 and NB3.
\label{fig:CO_shell}}
\end{figure}

With ALMA, CO emission can now be mapped at higher angular resolution and
sensitivity.  Figure \ref{fig:CO_chan} shows the CO maps from blueshifted to
redshifted velocities (with a velocity interval of 4.23 \vkm{}) in
comparison to the \H2{} map.  The blueshifted emission is mainly to the
north and redshifted mainly to the south of the central source.  Nested
shell structures are seen extending to the north and south from the central
source.  As shown in Figure \ref{fig:CO_shell}, the nested shell structures
can be better seen in the CO maps integrated over selected velocity ranges
that show the shell structures.  In the south (Figure \ref{fig:CO_shell}a),
a wide-opening shell is seen opening to the south from the central source,
connecting to the big bow shock SB3 further out.  Likewise in the north
(Figure \ref{fig:CO_shell}b), a similar wide-opening shell is seen opening
to the north, connecting to the big bow shock NB3 further out.  These
wide-opening shells are also seen in the lower transition line of CO at
J=2-1 \citep{Lee2006}, likely tracing the  outflow cavity walls
consisting mainly of the ambient material swept up by the big bow shocks. 
At the base around the central source (see the zoom-in in Figure
\ref{fig:CO_chanzoom}c), the shells fit right into the bay of the flattened
envelope detected in 350 GHz continuum \citep{Lee2014}.  This morphological
relationship suggests that the shells, as they expand, are sweeping up and
thus excavating the flattened envelope at the base.

Inside these wide-opening shells, internal shells are seen connecting to bow
shocks NB1/2 and SB1/2 from the central source (Figure \ref{fig:CO_shell}). 
Here internal shells are referred as the shells located inside the cavity
walls \citep{Lee2001}.  In the north, the internal shell is also affected by
the wings of bow shocks NK7 and NK8 inside the shell (Figure
\ref{fig:CO_chan}c-g).  Inside these shells, CO emission is also seen
surrounding other smaller knots closer to the source, forming narrow
internal shells around them.  These inner shells are better seen as we zoom
in to the inner part at low velocities as shown in Figure
\ref{fig:CO_chanzoom}.  For the prominent pair of knots, NK1 and SK1, where
the shocks are strong, hollow internal shells are clearly seen connecting to
them (Figs \ref{fig:CO_chanzoom}b and d), as seen earlier in
\citet{Lee2007}.  In summary, nested shells are seen in CO connecting to the
bow shocks and knots, driven by them.

\subsection{Molecular Jet in SiO and CO} \label{sec:jet}

Figure \ref{fig:jet} shows the maps of the SiO emission and high-velocity
(HV) CO emission in comparison to the \H2{} map.  As compared to the
previous SMA maps in \citet{Lee2007,Lee2008}, the maps here are created
using slightly larger velocity ranges, because ALMA detected the emission in
slightly larger velocity ranges at higher sensitivity.  For the SiO map, we
use the velocity from $-$23.03 to 17.53 \vkm{}, instead of $-$21.1 to 16
\vkm{}.  For the CO map, low-velocity emission is excluded to avoid the
shell contamination.  On the blueshifted side, we use the velocity from
$-$21.65 to $-$8.95 \vkm{}, instead of $-$18.4 to $-$7.1 \vkm{}.  On the
redshifted side, we use 9.56 to 18.77 \vkm{}, instead of 3.4 to 13.3 \vkm{}. 
Notice that the high-velocity ranges are not symmetric with respect to the
systemic velocity because the jet has a higher mean velocity and a larger
velocity range on the blueshifted side than the redshifted side \cite[see
Figure 5 in][]{Lee2007}, probably due to different inclination angles of the
jet in the northern and southern parts {\bf \citep{Lee2007}}. As found in
\citet{Lee2007,Lee2008}, both SiO emission and HV CO emissions trace the jet
well, arising from the sinuous structures and knots in the jet seen in
H$_2$.  Notice that since the H$_2$ map was made earlier, the SiO and CO
emission peaks are seen slightly ahead of the H$_2$ emission peaks.  As
discussed in \citet{Lee2007}, the jet has a slight bending of $\sim$
1.5\degree{} to the west and a small semi-periodical wiggle in the
trajectory.

Due to higher sensitivity of ALMA, SiO emission becomes better
detected toward the central source position, as compared to that seen in
\citet{Lee2008}, likely arising from the jet base near the source as seen in
the lower transition line of SiO at J=5-4 \citep{Codella2007}.  Also,
the jet appears to be more continuous in CO, especially in the northern
part.  In addition, three more knots are detected along the jet axis in the
south, where no clear H$_2$ knot was detected.  One is detected in SiO and
CO in between knots SK5 and SK7, and is thus labeled as knot SK6 as a
counterpart of knot NK6 in the north.  Two are detected in CO in between
knots SK6 and SK7, linking the two knots. The \H2{} and CO emission seen in
between NK6 and NK7 could be their counterparts in the north.  For the
prominent knot SK1 where the shock is strong, the SiO emission is seen only
at the bow tip, while the CO emission is seen mainly in the bow shock wings. 
The limb-brightened bow wings structure of the CO emission can be seen at
lower velocity (Figure \ref{fig:CO_chanzoom}d).  Similar morphology is also
seen in L 1157 in \citet{Gueth1998}, probably because CO is destroyed or
excited to higher transition lines at the tip where the shock is stronger.

The jet can also be seen at lower velocity (see Figure \ref{fig:CO_chan})
because of the low inclination angle of the jet to the plane of the sky. 
Interestingly, at higher angular resolution and sensitivity, the CO
emission at 5.86 \vkm{} appears to show a helical structure along the jet
axis in the south (see Figure \ref{fig:CO_chanzoom}d) that is not seen
before. This helical structure is actually consisted of a chain of small
and similar-size bow shocks curving back to the jet axis, associated with
the knots (SK2, SK3, SK4, and SK5) in the jet.  The bow wings are faint in
the west, causing them to appear as a helical structure.  In addition, a
faint jetlike structure can also be seen along the jet axis in between knots
SK3 and SK4 at this velocity, tracing the jet itself.  SiO emission also
shows the similar bow shocks curving back to the jet axis (see Figure
\ref{fig:jet}b).  This feature has been seen in the simulations of a pulsed
jet, both unmagnetized \citep{Stone1993,Biro1994,Volker1999} and magnetized
\citep{Stone2000}.  On the other hand, a pulsed wide-angle wind would
produce a chain of wide bow shocks that do not curve back to the jet axis
\citep{Lee2001}, inconsistent with our observations.

\subsection{Proper Motion}


Since the knots in the jet are well traced by the SiO emission, proper
motion of the jet can be estimated by measuring the position shifts of their
SiO emission peaks with respect to those seen at $\sim$ 6 years earlier in
our previous SMA map in \citet{Lee2008}.  We first convolved our previous
SMA SiO map to the resolution of our ALMA map and then aligned the two maps
with the continuum peaks.  As shown in Figure \ref{fig:proper}, the jet
pattern is similar in the two epochs, in agreement with the pattern being
propagating with the jet velocity.  Only the knots that are resolved and
have well defined peaks are used for the measurement, with their emission
peak positions marked with the solid lines.  The innermost pair of knots, SS
and SN, are not used for the measurement because they are spatially
unresolved in our ALMA observations, appearing as jetlike structures.  Thus,
the mean position shift is estimated to be $\sim$
\arcsa{0}{36}$\pm$\arcsa{0}{15}.  Hence, the proper motion is estimated to
be $\sim$ \arcsa{0}{06}$\pm$\arcsa{0}{025} per year, giving rise to a
tangential velocity of $\sim$ 115$\pm50$ \vkm{}.  Since the jet is almost in
the plane of the sky, the jet velocity can be approximated to this
tangential velocity.



\subsection{Jet Density and Mass-Loss Rate} \label{sec:den}

With better estimated jet velocity and better resolved CO emission than
those reported before in \citet{Lee2007}, we can refine the
jet density and thus the mass-loss rate.  The CO emission of the jet in
between knot NK1 to knot NK5 appears continuous and smooth (see Figure
\ref{fig:jet}c), and thus can be used to derive the mean jet density.  The
mean CO intensity there is found to be $\sim$ 1.2 \Jyb{} \vkm{}.  As
discussed in \citet{Lee2007}, the mean excitation temperature of the CO
emission can be assumed to be $\sim$ 50 K.  Since the brightness temperature
in the jet is mostly $\lesssim$ 20 K, the CO emission can be assumed to be
optically thin.  Assuming LTE, the column density of CO is estimated to be
$\sim 2.3 \times 10^{16}$ \cms.  The mean jet (\H2{} volume) density can be
derived from the CO column density using the conversion equation, e.g., Eq. 
17 in \citet{Lee2014}.  With a CO abundance of $8.5 \times 10^{-5}$ and a
jet diameter of $\sim$ \arcsa{0}{2} \citep{Cabrit2007,Lee2008}, the mean jet
density is estimated to be $\sim 5.6 \times 10^5$ \cmc{}, similar to the
mean value derived from \HCOP{} in \citet{Lee2014}.  Thus, the (two-sided)
mass-loss rate would be $\dot{M}_j\sim 1.1\times10^{-6}$ \solarmass{}
yr$^{-1}$.  Since the accretion rate has been estimated to be $\sim
5\times10^{-6}$ \solarmass{} yr$^{-1}$ \citep{Lee2014}, the mass-loss rate
is estimated to be $\sim$ 20\% of the accretion rate, reasonably
consistent with the 
X-wind model \cite[$\sim$ 30\%,][]{Shu2000} and the disk-wind model
\cite[$\sim$ 10\%,][]{Konigl2000}.





\section{Discussion}

\subsection{Bow Shock Formation and Sideways Ejection} \label{sec:bow}


As discussed earlier, the knots in the southern jet are seen with small bow
shocks, with their bow wings curving back to the jet axis.  Knots SK4 and
SK5 are reasonably resolved, allowing us to study the formation mechanism of
the bow shocks.   For knot SK5, the CO emission peak appears clearly
upstream of (i.e., closer to the central source than) the SiO emission peak
(Figure \ref{fig:SK4_5}a).  For knot SK4, the CO emission peak appears only
slightly upstream of the SiO emission peak by {\bf \arcsa{0}{08} or}
$\sim$ 15\% of the beam size,
and thus higher resolution observations are needed to confirm it. In SiO,
the position-velocity (PV) diagram cut along the jet axis shows an arclike
PV structure for each knot, with an apex at the velocity of $\sim$ 9 \vkm{}
pointing away from the source, as shown in Figure \ref{fig:SK4_5}b.  As we
move away from the apex, either to higher velocity or to lower velocity, the
PV structure bends slightly toward the source direction, forming the arclike
PV structure.  The CO emission shows a similar PV structure slightly
upstream of that of the SiO emission.  Note that the CO emission is missing
from $\sim$ 8 to 10 \vkm{} due to a foreground cloud at those velocities. 
In the northern jet, the knots are not spatially resolved and thus no
clear bow-like structures can be seen in SiO and CO.  Higher resolution
observations are needed to study them.

Arclike PV structure has been seen in pulsed jet simulations \cite[see e.g.,
Figure 16a in][]{Stone1993}.  In those simulations, even though the jet
source is continuously ejecting material with a constant density, a
(sinusoidal) variation in the ejection velocity produces an internal working
surface (IWS) in the jet \citep{Biro1994}.   The IWS can appear as a
small arclike knot, with its velocity increasing toward its two edges from
the jet axis \cite[see Figure \ref{fig:IWS} and also Figure 10
in][]{Lee2004}, producing the arclike PV structure. As its shocked material
is ejected sideways into the outflow cavity, it forms a bow shock, with the
bow wings curving back to the jet axis \citep{Biro1994}, like those of knots
SK4 and SK5 (see Section \ref{sec:jet}).  Therefore, knots SK4 and SK5
likely trace the IWSs in the jet.  A bigger variation in the jet velocity
would produce faster and more massive sideways ejection, forming stronger
bow shocks \citep{Raga2002}, such as SK1, NK1, NB1/2 and SB1/2, and thus
bigger internal shells extending backward to the central source from their
bow wings, as seen in the pulsed jet simulations in \citet{Biro1994} and
\citet{Lee2001}.

Sideways ejection has been claimed in another jet source IRAS 04166+2706
\citep{Santiago2009}.  In that jet, a sawtooth pattern was seen in the PV
diagram cut along the jet axis in CO and SiO, and it was argued to be
produced by the sideways ejection of the shocked material in the IWSs.  That
jet is highly inclined with an inclination angle of $\sim$ 45\degree{} to
the plane of the sky.  Viewing the IWSs at high inclination angle produces
the sawtooth PV pattern, as predicted in the simulated PV diagram at high
inclination angle in \citet{Stone1993}.  Here the HH 212 jet is almost in
the plane of the sky with a small inclination angle, thus the sideways
ejection forms the arclike PV structures in the PV diagram.

\subsection{Wide-Angle Wind Component?}

In addition to a collimated jet, an unseen wide-angle wind seems to be
needed in driving molecular outflows in the later stage of star formation
\citep{Lee2005,Arce2013}.   Is a wide-angle wind also needed in the
early phase of star formation as in HH 212? Recently, a wide-angle
flow has been seen in C$^{34}$S and may suggest a presence of a wide-angle
wind \citep{Codella2014}.  This wide-angle flow has also been seen earlier
in \HCOP{} \citep{Lee2014} and it may trace the cavity walls instead.  In
addition, a wide-angle wind, if exists, would produce wide bow-shocks and
thus wide internal shells \citep{Lee2001}.  This is inconsistent with our
observations, which show highly curved bow shocks and narrow shells driven
by the knots.  Therefore, in this source the wide-angle wind, if exists, is
not significant enough to affect the structures of the bow shocks and the
shells. This is not inconsistent with current jet launching models
\citep{Shu2000,Konigl2000}, which predict a wide-angle wind much more
tenuous and thus less significant than the central jet.  In the later phase,
the jet is much tenuous, so that the contribution of the wide-angle wind
could become more important \citep{Lee2005,Arce2013}.

\subsection{Jet Wiggle}

The jet wiggle can be better seen after combining the CO and SiO maps of
ALMA with the \H2{} map, as shown in Figure \ref{fig:wiggle}.  The northern
jet is continuous and the wiggle is reasonably well defined.  The southern
jet appears to be continuous only up to knot SK2 at $\sim$ $-$\arcs{10}, and
thus the wiggle is not so well defined.  Comparing the sinuous structures in
between knots SK1 and SK2 to those in between knots NK1 and NK2, the wiggle
in the southern jet could be roughly point-symmetric to that in the northern
jet.  Therefore, we first model the wiggle in the northern jet and then
apply the model to the southern jet assuming a point-symmetric wiggle about
the central source.


A possible mechanism to produce a point-symmetric wiggle is the jet
precession \citep{Raga2009}. In this case, the amplitude of the wiggle, $A$, will
increase linearly with the distance, i.e.,
\begin{equation} 
A (z)  = z \tan \theta_0
\end{equation} 
where $z$ measures the distance of the jet from the source along the jet axis. 
Here $\theta_0$ is the half-opening angle of the wiggle, so that
$\tan \theta_0$ gives the rate of the growth in the amplitude with the distance.
The wiggle appears to be semi-periodical and thus can be assumed to be given by the
following sinusoidal model
\begin{equation} 
x (z) = A (z) \sin (\frac{2 \pi z}{\lambda} + \phi_0) 
\end{equation}
where $x$ measures the displacement of the jet perpendicular from the jet
axis.  $\lambda$ is the wavelength of the wiggle and $\phi_0$ is the phase
angle at the source.  Based on the wiggle of the northern jet, the best
parameters are $\theta_0 \sim$ 0.5\degree{}$\pm$0.2\degree{}, $\lambda
\sim$ \arcsa{5}{6}$\pm$\arcsa{1}{0} (or 2240$\pm$400 AU), and $\phi_0 \sim$
0\degree{}$\pm$50\degree{}.  Since the jet velocity is $\sim$ 115 \vkm{},
the period of the wiggle is $\sim$ 93 years.  As seen in Figure
\ref{fig:wiggle}a, the model can roughly reproduce the wiggle of both the
northern and southern jet in the inner part, but it seems to predict a
larger wiggle than the observed in the outer part with $|z| \gtrsim$
\arcs{20}.

In order to improve the model, the amplitude of the wiggle is assumed to
first increase linearly with the distance and then remain constant at $z\geq z_0$, i.e.,
\begin{eqnarray}
A(z) = \left\{ \begin{array}{cl}
z \tan \theta_0 & \;\;\textrm{if}\;\; z < z_0,  \\ 
z_0 \tan \theta_0 & \;\;\textrm{if}\;\; z \geq z_0
\end{array}  \right.
\end{eqnarray}
As shown in Figure \ref{fig:wiggle}b, the model with $z_0 \sim$ \arcs{10}
(4000 AU) can give a better match up to $\sim$ \arcs{34} (i.e., 6 cycles of
wiggle) in the northern jet and up to $\sim$ \arcs{38} in the southern jet,
where the SiO and CO emission were detected.  At $z = z_0$, the amplitude of
the wiggle reaches the maximum value of $A_m \equiv z_0 \tan \theta_0 \sim
\arcsaq{0}{1}$, roughly the same as the jet radius, which is also $\sim$
\arcsa{0}{1} \citep{Cabrit2007,Lee2008}.  Note that, since the southern jet
does not have continuous structure beyond $\sim$ $-$\arcs{10} from the
source, further observations are still needed to confirm that the wiggle of
the jet is really point symmetric about the central source.  Observations at
higher angular resolution are also needed to refine the value of $z_0$.



If the wiggle is really point-symmetric about the central source, then one
way to produce the wiggle is the jet precession due a tidal interaction of a
binary companion on the disk of the jet source \citep{Terquem1999}. 
However, if this is the case, the amplitude of the wiggle is expected to
grow with the distance \citep{Raga2009}, inconsistent with our observations.




Another possibility is the current-driven kink instability
\citep{Cerqueira2001} since the jet is expected to be magnetized with a
helical magnetic field morphology \citep{Shu2000,Konigl2000}.  In this case,
the amplitude of the wiggle is expected to grow at the beginning, and then
become saturated and thus remain roughly constant at large distance, see,
e.g, Figure 3 in both \citet{Cerqueira2001} and \citet{Mizuno2014}.  In
addition, the wiggle is more like oscillation and thus could be
semi-periodical.  Both of these are roughly consistent with what we observe
here in our jet.  However, the wiggle is not expected to show any symmetry
about the central source.  This might be OK because the jet wiggle here
might not be point-symmetric, as mentioned earlier.  In addition, the
detailed jet structure here already appears to be different between the
northern jet and southern jet.  For the kink instability to take place, we
have the Kruskal-Shafranov criterion $|B_p/B_\phi| < \lambda/ 2 \pi A_m$
\cite[e.g.,][]{Bateman1978}.  With the maximum displacement $A_m \sim$
\arcsa{0}{1} and the wiggle wavelength $\lambda \sim$ \arcsa{5}{6}, we have
$|B_p/B_\phi| < 9$.  Therefore, this kink instability, if in action, could
be initiated in the central part of the jet, where the magnetic field is
dominated by the poloidal field \citep{Pudritz2012}.  This poloidal field
could serve a ``backbone" to stabilize the jet \citep{Ouyed2003}.  The
toroidal field dominates only near the jet edges in order to collimated the
jet.

\section{Conclusions}

We have mapped the HH 212 jet and outflow system in the SiO (J=8-7) and CO
(J=3-2) lines. Our primary conclusions are the following:

\begin{itemize} 

\item Although the jet is seen with knots and bow shocks, the underlying jet
is continuous, especially in the northern side.

\item The proper motion of the jet is estimated to be $\sim$ 115$\pm$50
\vkm{}.  The mass-loss rate in the jet is refined to be $\sim$
$1.1\times10^{-6}$ \solarmass{} yr$^{-1}$, as expected for a Class 0 source.

\item Some knots are seen associated with small bow shocks, with their bow
wings curving back to the jet axis.  Two of them, e.g., knots SK4 and SK5,
are reasonably resolved, showing kinematics consistent with sideways
ejection, likely tracing the internal working surfaces formed in a
collimated jet due to a temporal variation in the jet velocity.  Their bow
shocks are formed by this sideways ejection.

\item The jet has a small semi-periodical wiggle, with a period of $\sim$ 93
yrs.  The amplitude of the wiggle first increases with the distance and then
stays roughly constant.  A jet precession may have difficulty to explain this
kind of wiggle.  One possible origin of the wiggle could be the kink
instability in a magnetized jet.

\item  Nested (internal) shells are seen in CO connecting to the knots and
bow shocks, driven by them.  The sideways ejection ejects material into the
outflow cavity from the knots, forming the bow shocks and the internal
shells.

\item The internal shells of the knots are narrow and the bow shocks of the
knots curve back to the jet axis, both suggesting that the wide-angle wind,
if exists, is insignificant as compared to the jet.

\end{itemize}

\acknowledgements

We thank the anonymous referee for insightful comments.
This paper makes use of the following ALMA data:
ADS/JAO.ALMA\#2011.0.00647.S.  ALMA is a partnership of ESO (representing
its member states), NSF (USA) and NINS (Japan), together with NRC (Canada)
and NSC and ASIAA (Taiwan), in cooperation with the Republic of Chile.  The
Joint ALMA Observatory is operated by ESO, AUI/NRAO and NAOJ.  These data
were made available to Chin-Fei Lee, as part of his ALMA proposal
2011.0.00122.S (PI: Chin-Fei Lee), which requested observations duplicating
those of proposal 2011.0.00647.S.  C.-F.  Lee acknowledges grants from the
National Science Council of Taiwan (NSC 101-2119-M-001-002-MY3) and the
Academia Sinica (Career Development Award).





\begin{figure} [!hbp]
\centering
\putfiga{1.2}{270}{f3.ps} 
\figcaption[]
{CO channel maps (Red, in linear scale) in comparison with the \H2{} map of
the HH 212 system in the inner part.  The synthesized beam in the CO maps
has a size of \arcsa{0}{56}$\times$\arcsa{0}{47}, as indicated in the lower
right corner in \tlabel{a}.  The star marks the central source position. 
The channel velocity is indicated in the lower left corner.  The contours
show the continuum map at 350 GHz adopted from \citet{Lee2014}.
\label{fig:CO_chanzoom}}
\end{figure}

\begin{figure} [!hbp]
\centering
\putfiga{0.95}{270}{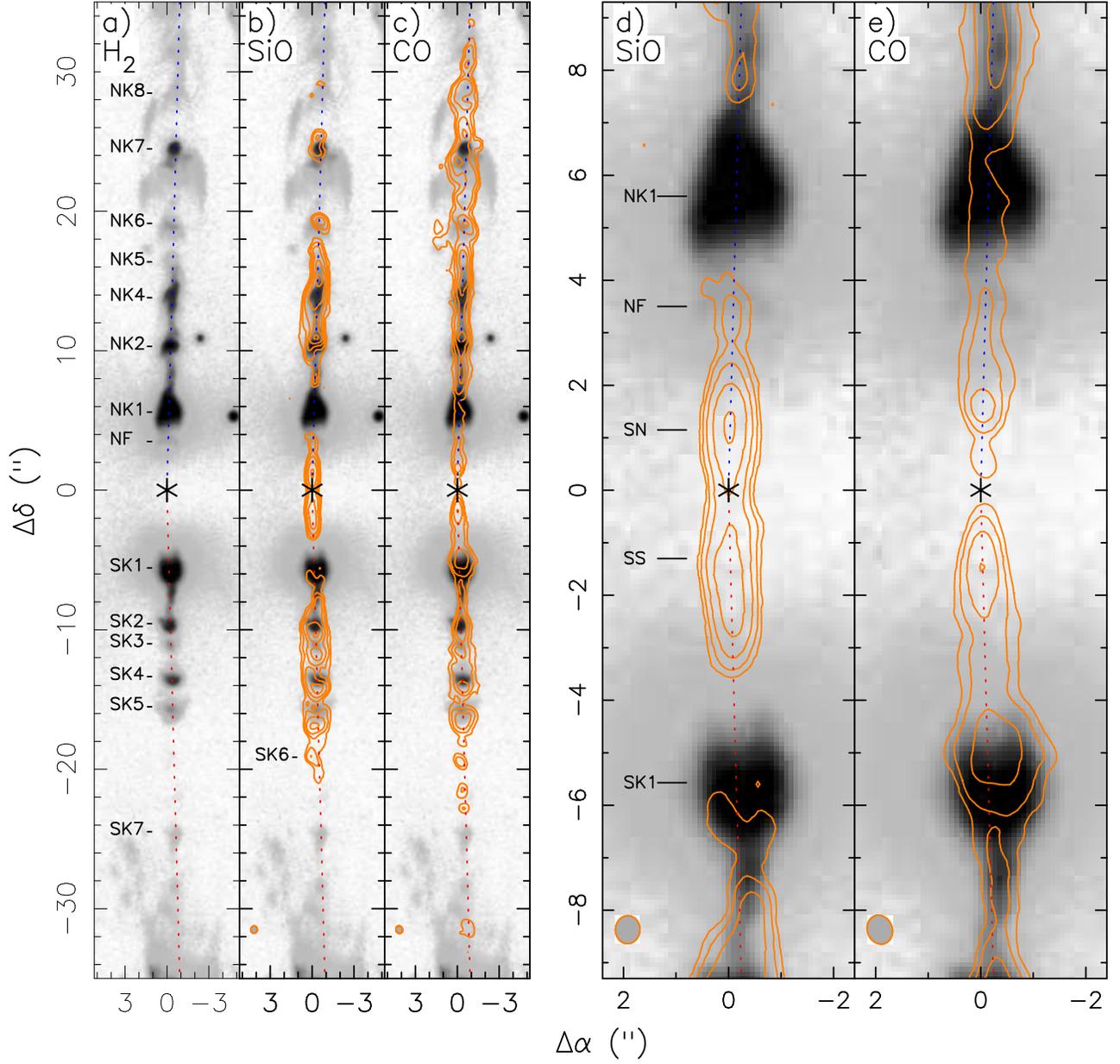} 
\figcaption[]
{The jet in SiO, CO, and \H2{}. Gray image shows the \H2{} map.
In \tlabel{b}, contours show the SiO map (integrated from
$-$23.03 to 17.53 \vkm{}).  Contour levels are $0.23 \times 2.5^{n-1}$ \Jybk,
where $n=1,2,3..$.  In \tlabel{c}, contours show the high-velocity CO map
(integrated from $-$21.65 to $-$8.95 \vkm{} on the blueshifted side and from
9.56 to 18.77 \vkm{} on the redshifted side).  Contour levels are
$0.20 \times 2^{n-1}$ \Jybk, where $n=1,2,3..$.  Panels \tlabel{d} and
\tlabel{e} are the zoom-in of panels \tlabel{b} and \tlabel{c},
respectively.
\label{fig:jet}}
\end{figure}

\leftblank{
\begin{figure} [!hbp]
\centering
\putfiga{1}{270}{f5.ps} 
\figcaption[]
{CO channel map at 5.86 \vkm{} zooming in to the inner part in the south.
It is the same as Figure \ref{fig:CO_chanzoom}d but without the \H2{} map.
The color scheme is different in order to highlight
the small bowlike structures curving back to the jet axis.
\label{fig:COhel}}
\end{figure}
}

\begin{figure} [!hbp]
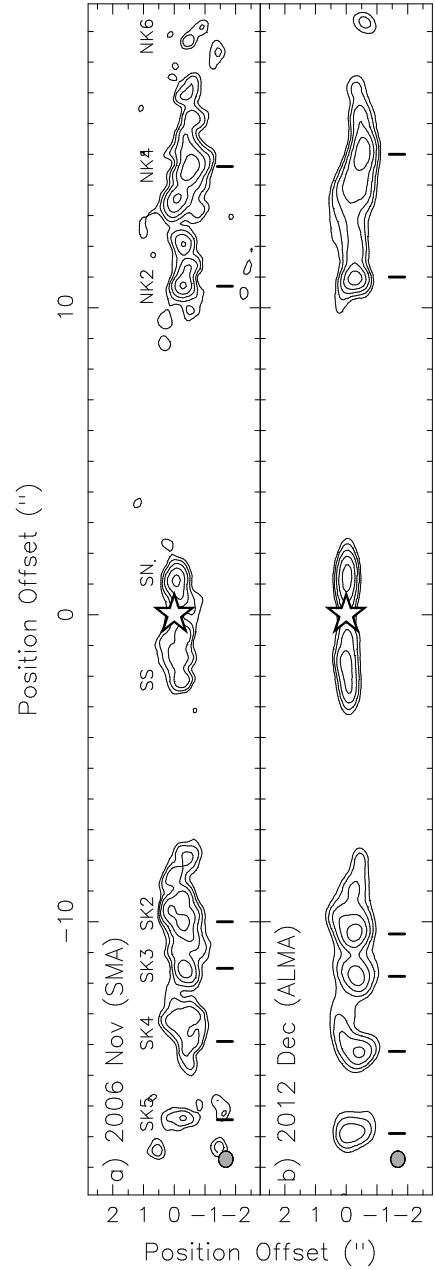

\centering
\putfiga{1}{270}{f5.ps} 
\figcaption[]
{Proper motion measurement using the SiO knots in the jet in two different
epochs at the ALMA angular resolution of \arcsa{0}{56}$\times$\arcsa{0}{47}. 
Contour levels are $1.2\times1.6^{n-1}$ \Jybk, where $n=1,2,3..$.  The star
marks the central source position.  The line segments indicate the SiO peak
positions of the knots.
\label{fig:proper}}
\end{figure}

\begin{figure} [!hbp]
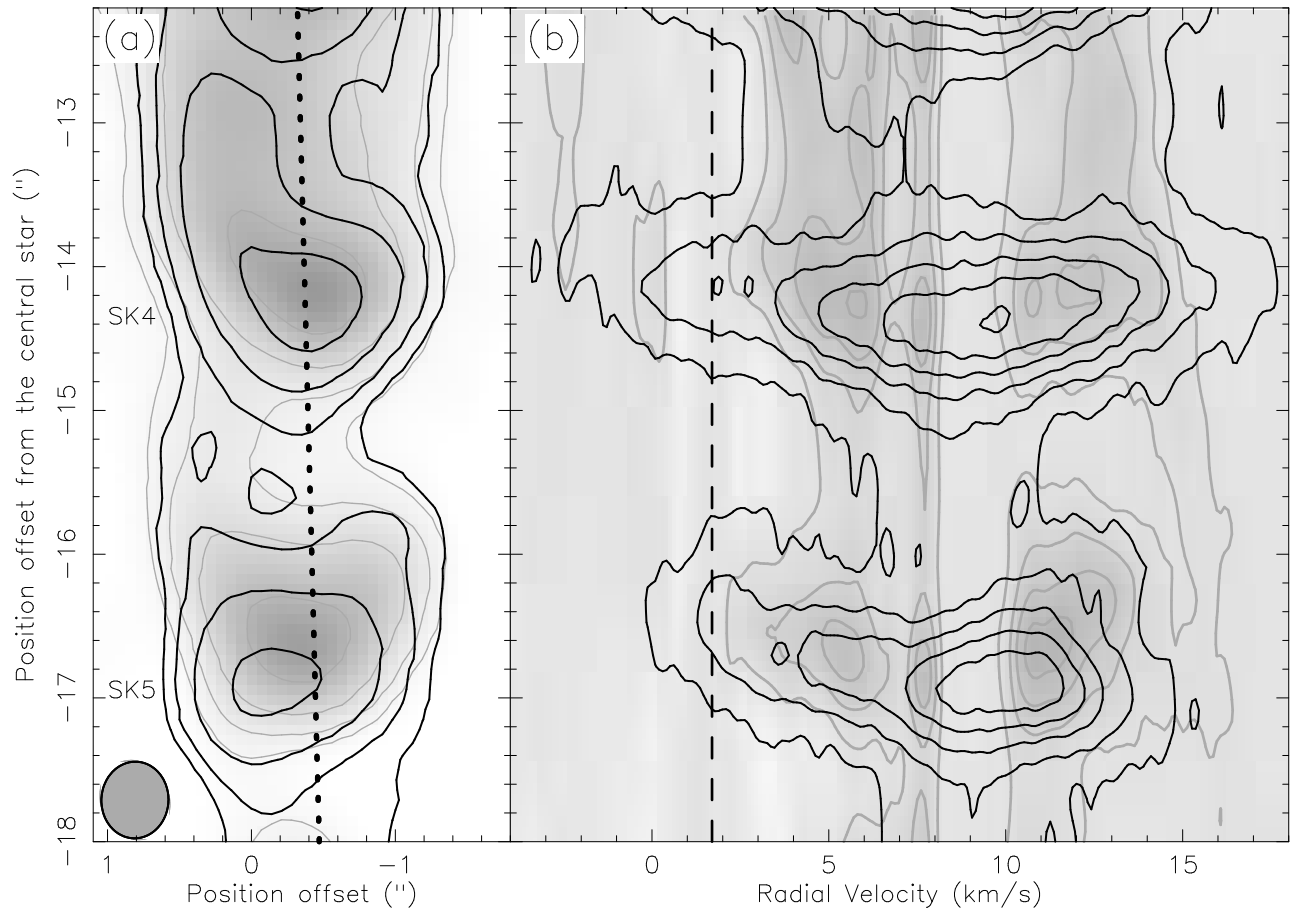

\centering
\putfiga{0.7}{270}{f6.ps} 
\figcaption[]
{Structure and kinematics of knots SK4 and SK5 in the south.  Gray contours
with image are for CO and black contours are for SiO.  \tlabel{a} The
SiO map is the same as in Figure \ref{fig:jet} but zooming into
knots SK4 and SK5, highlighting their bowlike structures.  The CO map shows the
total CO emission of knots SK4 and SK5 integrated from 2.2 to 15.4 \vkm{}. 
Contour levels are $0.25 \times 2^{n-1}$ \Jybk, where $n=1,2,3..$. The
dotted line indicates the jet axis in the south.  \tlabel{b}
Position-velocity (PV) diagrams cut along the jet axis for knots SK4 and
SK5.  The vertical dashed line indicates the systemic velocity of 1.7
\vkm{}.  The contour levels start at 40 \mJyb{} with a step of 160 \mJyb{}
for CO, and start at 30 \mJyb{} with a step of 120 \mJyb{} for SiO.
\label{fig:SK4_5}}
\end{figure}

\begin{figure} [!hbp]
\centering
\putfiga{0.5}{270}{f7.ps} 
\figcaption[]
{Velocity structure of the shocked material in an internal working surface
(IWS) in the shock frame moving down the jet axis.  The vectors show the
material motion, with the length showing the magnitude.  Two layers, CO and
SiO, are plotted as an illustration based on our observational results.
\label{fig:IWS}}
\end{figure}

\begin{figure} [!hbp]
\centering
\putfiga{1.2}{270}{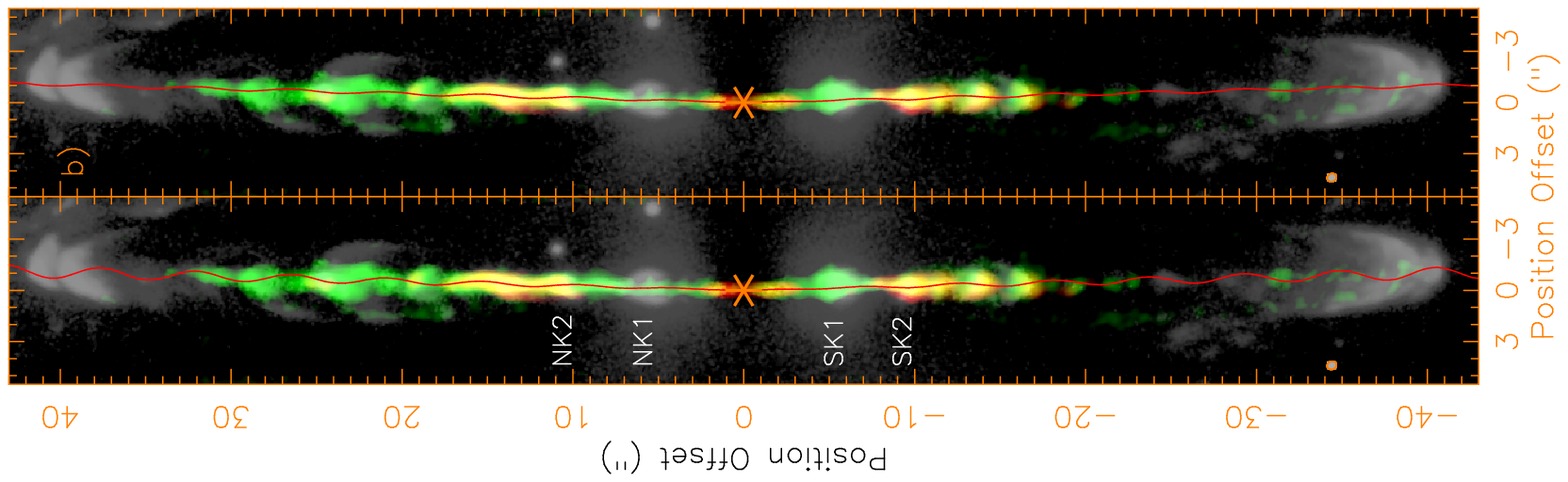} 
\figcaption[]
{The wiggle of the jet and the model (red curves) for the wiggle. Gray image
is \H2{}, green image is high-velocity CO, and red image is SiO, all from
Figure \ref{fig:jet}.  \tlabel{a} The model assumes a constant growth in the
wiggle amplitude with the distance from the central source.  \tlabel{b} A
revised model assumes a constant wiggle amplitude for a distance greater
than \arcs{10} from the central source.
\label{fig:wiggle}}
\end{figure}

\end{document}